\documentclass[12pt]{iopart}

\usepackage[pdftex]{graphicx}
\usepackage{iopams}
\usepackage{epstopdf}
\usepackage{cite}
\usepackage{gensymb}

\begin{document}
\def\be{\begin{equation}}
\def\ee{\end{equation}}
\def\bea{\begin{eqnarray}}
\def\eea{\end{eqnarray}}
\def\rp{r_{+}}
\def\rmm{r_{-}}

\title{Potts Models with (17) Invisible States on Thin Graphs}

\date{March 2013}
\author{D. A. Johnston}
\address{Dept. of Mathematics and the Maxwell Institute for Mathematical
Sciences, Heriot-Watt University,
Riccarton, Edinburgh, EH14 4AS, Scotland}

\author{R. P. K. C. M. Ranasinghe}
\address{Department of Mathematics, University of Sri Jayewardenepura,
Gangodawila, Sri Lanka.}


\begin{abstract}
The order of a phase transition is usually determined by the nature of the symmetry breaking at the phase transition point and the dimension of the model under consideration. For instance, $q$-state Potts models in two dimensions display a  second order, continuous transition for $q = 2,3,4$ and first order for higher $q$. 

Tamura {\it et al} recently introduced Potts models with ``invisible'' states which contribute to the entropy but  not the internal energy and noted that adding such invisible states could transmute  continuous transitions into first
order transitions \cite{TT1,TT2,TT3,TT4}. This  was observed both in a Bragg-Williams type mean-field calculation and $2D$ Monte-Carlo simulations.
It was suggested that the invisible state mechanism for transmuting the order of a transition might play a role where transition orders inconsistent with the usual scheme had been observed.

In this paper we note that an alternative mean-field approach employing 
$3$-regular random (``thin'') graphs also displays this change in the order of the transition as the number of invisible states is varied, although the number of states required to effect the transmutation, $17$ invisible states when there are $2$ visible states, is much higher than in the Bragg-Williams case. The calculation proceeds by using the equivalence of the Potts model with $2$ visible and $r$ invisible states to the Blume-Emery-Griffiths (BEG) model, so a by-product is the solution of the BEG model on thin random graphs.

\end{abstract} 

\maketitle


\section{Introduction}
The idea that spontaneous breaking of symmetries gives rise to phase transitions plays an important role in our understanding of phase transitions in statistical physics, particle physics and even computer science. A canonical model for understanding the breaking of $q$-fold symmetries is the
$q$-state Potts model \cite{Potts}, whose Hamiltonian for uniform ferromagnetic couplings may be written as
\begin{equation}
\label{Potts1}
 \, {\cal H}_q =   -   \sum_{\langle ij \rangle}  \delta_{\sigma_i, \sigma_j}
\end{equation}
where the interactions  are along the edges $\langle ij \rangle$ of some graph, the spins $\sigma_i$
at each vertex may take $q$ values and $\delta$ is the Kronecker delta. With this Hamiltonian like spins interact with one  strength and unlike spins with a different strength.  As is well known the phase transition for the Potts model in two dimensions is continuous, i.e. second order, if $q=2,3,4$
and first order for larger $q$. It is generally expected that a phase transition associated with the spontaneous breaking of a $q$-fold symmetry in two dimensions will have the same order as the ferromagnetic  $q$-state Potts model, but recent studies have suggested this is by no means always the case \cite{q3,q31,q32}.

It was observed in \cite{TT1,TT2,TT3,TT4} that one possible mechanism giving rise to  such 
non-canonical orders for phase transitions might be the presence of
``invisible'' states which did not contribute to the internal energy but 
still contributed to the entropy. It is possible to write down a Potts
Hamiltonian that incorporates such states as
\begin{eqnarray}
 \label{eq:original_Hamiltonian}
  {\cal H}_{(q,r)} = -  \sum_{\langle i,j \rangle}
  \delta_{s_i, s_j} 
  \sum_{\alpha = 1}^{q} \delta_{s_i, \alpha} \delta_{s_j, \alpha},
  \,\,\,\,\,\,\,
  s_i = 1, \cdots, q, q + 1, \cdots, q + r,
\end{eqnarray}
 which has $q$  visible states and $r$  invisible states
 for each spin $s_i$. In the sequel we will denote this as the ($q$,$r$)-state Potts model.
 
 A useful way to rewrite this Hamiltonian is to introduce
 new spin variables $\sigma_i$ where
 $\sigma_i = s_i$ if $s_i =1, \cdots, q$ and $\sigma_i=0$ otherwise \cite{TT1}. The
resulting Hamiltonian 
\begin{eqnarray}
 {\cal H}'_{(q,r)} =  -  \sum_{\langle i,j \rangle} \delta_{\sigma_i, \sigma_j}
\sum_{\alpha = 1}^q \delta_{\sigma_i, \alpha} \delta_{\sigma_j, \alpha}
 - T \ln r \sum_i \delta_{\sigma_i, 0},
 \,\,\,\,\,\,\,
 \sigma_i = 0,1,\cdots,q,
\end{eqnarray} 
 contains a temperature dependent field term
which gives the entropy contribution of the invisible states and the additional Kronecker deltas in the first interaction term ensure the invisible states do not contribute to the internal energy.
 
In the case of a Potts model with $2$ visible states and $r$ invisible states we may map this Hamiltonian in turn to that of a spin one model, the Blume-Emery-Griffiths (BEG) \cite{BEG} model with equal couplings
\begin{eqnarray}
\label{BEG}
 {\cal H}_{\rm BEG} &=& - {1\over 2} \sum_{\left\langle i,j \right\rangle}
  t_i t_j     - {1\over 2} \sum_{\left\langle i,j \right\rangle}    t_i^2 t_j^2 
 - \mu \sum_i \left( 1 - t_i^2 \right), \nonumber \\
  &t_i& = +1,\, 0,\, -1
\end{eqnarray}
where $\mu = T \ln r$. The generic BEG model admits non-equal couplings for the 
$t_i t_j$ and $t_i^2 t_j^2$ terms, but this is unnecessary for our purposes.
The  phase diagram for the BEG model has been studied by numerous authors using various means. In \cite{TT4}
a Bragg-Williams mean-field approximation was used, 
giving a phase diagram whose  general form is shown in Fig.\,(1) 
\begin{figure}[h]
\begin{center}
\includegraphics[height=5cm]{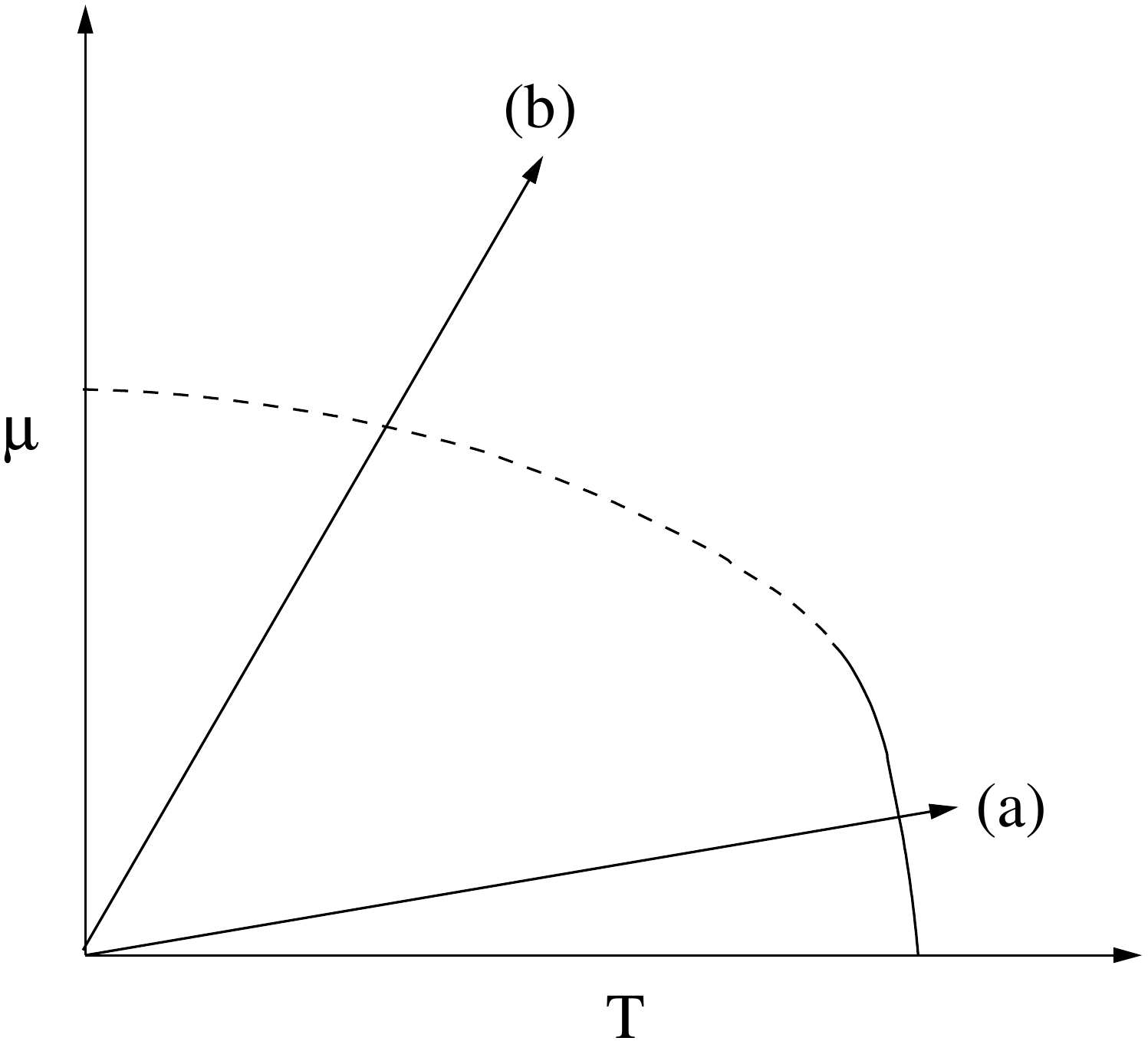}
\label{BEGgraph} 
\caption{The schematic drawing of the (equal coupling) mean-field BEG model phase diagram
in the $\mu,\, T$ plane. The second order transition line is shown in bold and the first order dashed. The arrowed sloping lines (a), (b) represent different values of $r$ when $\mu = T \ln r$ to make contact with the ($2$,$r$)-state Potts model. }
\end{center}
\end{figure} 
where a magnetized phase exists in the small $\mu,\, T$ region. The parameter $\mu$ is usually called the crystal field as a consequence of the model's phenomenological roots.
 
The mechanism by which the invisible states may affect the order of the transition is clear from the diagram. For smaller $r$ as $T$ is increased 
the system will move along a line similar to the the lower of the two sloping lines (since $\mu = T \ln r$), which we have denoted by (a) in Fig.\,(1). It will thus cross the second order
portion of the phase transition line. For sufficiently large $r$ the line is steep enough to cross the first order portion of the transition line, as shown in (b). Quantitatively, a second order transition occurs if $(q,r) = (2,1),\, (2,2), \, (2,3)$ and the transition becomes first order for larger $r$. The existence of the  low temperature symmetry broken phase and first order transition for the ($q$,$r$)-state model has also been shown rigorously using random cluster methods in \cite{AvE} for $q>1$ and sufficiently large $r$.

In this paper we employ another mean-field like model to study the $(2,r)$ state Potts model by making use of the fact that spin models on regular random graphs (random graphs where each vertex has the same valency) will generically
display mean-field  behaviour since such graphs look locally like a Bethe lattice \cite{Bethe}. Unlike the Bethe lattice regular random graphs {\it do} have large loops \cite{0b}, which renders them closed and obviates the need to deal with boundary conditions, as one must  for the Bethe lattice proper.
The weightings to use for spin models living on such graphs can be determined by using similar methods to those used for ``fat'' random graphs, where the perturbative expansion of matrix integrals over $N \times N$ matrices in the limit $N \to \infty$ give planar random graphs decorated with the appropriate Boltzmann weights. The regular random graphs considered here are generated in the  $N \to 1$, scalar limit of such integrals \cite{0,00}
so we denote them as ``thin'' random graphs. 

\section{Thin graphs}

The required ensemble of random graphs can be  generated from the Feynman diagram expansion of a scalar integral. For instance, if we are simply interested in calculating the number of undecorated 3-regular random graphs with $n$ vertices we could evaluate the integral
\begin{equation}
N_n = {1 \over 2 \pi i} \oint { d \lambda \over
\lambda^{2n + 1}} \int_{-\infty}^{\infty}{ d \phi \over \sqrt{2 \pi}} \; \exp \left( -\frac{1}{2} \phi^2 + \frac{\lambda}{6} \phi^3 \right)
\end{equation}
which generates the graphs in a perturbative expansion with a unit propagator from the $\frac{1}{2}\, \phi^2$ term tying together the cubic vertices from the 
$\frac{\lambda}{ 6} \, \phi^3$ term. The contour integral in $\lambda$ picks out the 
desired size of graph. Expanding the cubic terms in the exponential and evaluating the 
Gaussian integrals 
gives the  well known result
\begin{equation}
N_n = \left( {1 \over 6} \right)^{2n} { ( 6 n - 1 ) !! \over ( 2 n ) !
}.
\end{equation}
for the number of $3$-regular random graphs with $n$ vertices.

When the graphs are decorated with spins we introduce additional variables 
in the integral to give the desired weights to edge interactions. For the Ising model, for example, the  Hamiltonian on any graph is
\begin{equation}
{\cal H}_I = - \sum_{<ij>}  \sigma_i \sigma_j ,
\end{equation}
where the sum is over the edges connecting nearest neighbour sites.
The partition function
\begin{equation} 
Z_n( \beta)= \sum_{\{\sigma\}} \exp \left( \beta \sum_{\left<i j \right>} \sigma_i \sigma_j \right)
\end{equation} 
for the Ising model on an ensemble of thin graphs with $n$ vertices is then given by the integral \cite{0}
\begin{equation}
\label{ZnIsing}
Z_n(\beta) \times N_n = {1 \over 2 \pi i} \oint { d \lambda \over
\lambda^{2n + 1}} \int {d \phi_+ d \phi_- \over 2 \pi \sqrt{\det K}}
\exp (- S_I ),
\end{equation}
where the propagator $K$ is given by
\begin{equation}
\begin{array}{cc} K_{ab}^{-1} = & \left(\begin{array}{cc}
1 & -c \\
-c & 1
\end{array} \right) \end{array}
\end{equation}
with $c=\exp(-2 \beta)$
and the integrand, which we shall call an ``action'' since we are thinking
in terms of Feynman diagrams, is
\begin{equation}
S_I = {1 \over 2 } \sum_{a,b}  \phi_a  K^{-1}_{ab} \phi_b  -
{\lambda \over 3} (\phi_+^3 + \phi_-^3).
\label{e3}
\end{equation}
Inverting the propagator  gives the correct Ising interactions between the two
spin states, which are proxied by the two variables $\phi_+,\,\phi_-$ in the integral.
The approach extends naturally to other spin models, such as the q-state Potts models where
we may take an action of the form \cite{JP}
\begin{equation}
S_q = { 1 \over 2 } \sum_{i=1}^{q} \phi_i^2 - c \sum_{i<j} \phi_i \phi_j -{\lambda \over 3} \sum_{i=1}^q \phi_i^3
\label{qstate}
\end{equation}
with
\begin{equation}
c= { 1 \over \exp( 2 \beta ) + q-2}
\end{equation}
and calculate the equivalent of equ.\,(\ref{ZnIsing}) with the $q$ variables $\phi_i$. The Potts transition is found to be first order for $q\ge3$, as one might expect for a mean-field model.

To evaluate such spin model partition functions on thin graphs in the thermodynamic limit we can employ saddle point methods, in which case the vertex coupling $\lambda$ may be scaled out and we obtain the leading term by simply solving the saddle point equations for the action: $\partial S / \partial \phi_i =0$. Phase transitions occur when the free energies, given to leading order by $S$ itself, of the different solution branches cross over as the temperature or other parameters are varied. Using saddle point calculations for the thin graphs gives a straightforward approach to mean-field theory for spin models in general and it has been possible to show that the saddle point equations for  the thin graph actions may be transformed into the fixed point equations  of the recursion relations \cite{JP2} used to solve various spin models on Bethe lattices. 
 
\section{BEG on thin graphs}
The appropriate $S$ for the BEG model on thin (or indeed fat) graphs  is straightforward to write down \cite{3M1,3M2},
even for the more general case of a BEG model with
unequal couplings where the Hamiltonian is given by
\begin{eqnarray}
\label{BEG2}
 {\cal H}_{\rm BEG} &=& -{J \over 2} \sum_{\left\langle i,j \right\rangle}
  t_i t_j   - {K \over 2}  \sum_{\left\langle i,j \right\rangle}   t_i^2 t_j^2  
 - \mu \sum_i \left(1-  t_i^2 \right) \; .
\end{eqnarray}
Since this is  a spin one model, we employ three variables in $S_{\rm BEG}$
\begin{eqnarray}
S_{\rm BEG} &=& { 1 \over 2 } ( \phi_1^2 +\phi_2^2 + \phi_3^2)  - a (1-b) ( \phi_1 \phi_3 + \phi_2 \phi_3)
\nonumber \\
 &-&  b \, \phi_1 \phi_2  -  { 1 \over 3 } ( \phi_1^3 + \phi_2^3 + \Delta  \,  \phi_3^3)
\label{SBEG}
\end{eqnarray}
where the couplings in the action $S_{\rm BEG}$ are related to those in the Hamiltonian by
\begin{eqnarray}
\exp ( - \beta J ) &=& { b + a^2 ( 1-b)^2 \over 1 - a^2 ( 1-b)^2}\nonumber \\
\exp (- \beta K ) &=&    {(1-b^2)^2 a^4 \over \left[ 1 - a^2 (1-b)^2 \right] \left[ b + a^2 ( 1 - b)^2 \right]} \\
a^3 \, \exp (\beta \mu) &=& \Delta \nonumber
\label{BEGrels}
\end{eqnarray}
as can been seen by inverting the quadratic terms in $S_{\rm BEG}$ to obtain the propagator and demanding that the edge weights for the different spin configurations match those generated by the Hamiltonian of equ.\,(\ref{BEG2}).

For the purposes of investigating the $(2,r)$ state Potts model it is sufficient to 
take $J=K=1$, which means $S_{\rm BEG}$ simplifies to 
\begin{eqnarray}
S_{\rm BEG} &=& { 1 \over 2 } ( \phi_1^2 +\phi_2^2 + \phi_3^2)  - a ( \phi_1 \phi_3 + \phi_2 \phi_3) -  { 1 \over 3 } ( \phi_1^3 + \phi_2^3 + \Delta  \,  \phi_3^3) \; .
\label{SBEGeq}
\end{eqnarray}
Since $b=0$ when $J=K=1$  the relation between the Hamiltonian couplings and those in $S_{\rm BEG}$ also simplifies to 
\begin{eqnarray}
\label{BEGrels2}
\exp ( - \beta  ) &=& {a^2 \over 1 - a^2} \nonumber \\
a^3 \, \exp (\beta \mu) &=& \Delta  \; .
\end{eqnarray}
Inverting the first of these we obtain the relation between the coupling $a$ and $\beta$
\begin{equation}
a =\sqrt{{ 1 \over e^{\beta} + 1}}
\end{equation} 
so the physical range of $a$ is given by $0<a<1/\sqrt{2}$.
With the expression for $S_{\rm BEG}$ in hand, the saddle point equations for the model are given by
\begin{eqnarray}
{\partial S_{\rm BEG} \over \partial \phi_1} &=& \phi_1 -a \phi_3 - \phi_1^2   = 0 \nonumber \\
{\partial S_{\rm BEG} \over \partial \phi_2} &=& \phi_2 -a \phi_3 - \phi_2^2    =0 \\
{\partial S_{\rm BEG} \over \partial \phi_3} &=&  \phi_3 -a (\phi_1 + \phi_2)  - \Delta \phi_3^2  =0 \nonumber 
\end{eqnarray}
and these may then be used to obtain the phase diagram.
 
The propagator in $S_{\rm BEG}$ has a canonical form with 
unit coefficients for each of the quadratic terms in equ.(\ref{SBEG},\ref{SBEGeq}). This means that the $\phi_3^3$ vertices
must pick up an extra $a^3$ coefficient to give the correct Boltzmann weights for edges 
with $(0,0)$ spins.
This manifests itself in the relation between $\Delta$ and and the crystal field coupling $\mu$ in the original Hamiltonian in equs.(\ref{BEGrels},\ref{BEGrels2}). 
An alternative would be to  scale $\phi_3 \rightarrow \phi_3 / a$ to shift all this dependence to the propagator. 
We have belaboured this point because the relation between $\mu$ (in the BEG spin model Hamiltonian) and $\Delta$ (in the random graph action) 
determines
the number of invisible states $r$ via $\mu = T \ln r$ in equ.(\ref{BEGrels2})
\begin{equation}
\label{a3r}
a^3 r = \Delta
\end{equation} 
when we think of the BEG model as a representation of the ($2,r$)-state Potts model.

Perhaps the most direct way to see the phase structure of the model is to evaluate the magnetization
along appropriate slices of the two-dimensional phase diagram shown in Fig.\,(1). 
The natural choice of variables for $S_{\rm BEG}$ is given by $\Delta, a$, which we can translate back to $\mu, \beta$ as appropriate. The magnetization $M$ of a saddle point solution may be calculated by substituting the saddle point values
of $\phi_{1,2,3}$ into
\begin{equation}
M = {  \phi_1^3  -  \phi_2^3   \over  \phi_1^3 + \phi_2^3 + \Delta  \,  \phi_3^3 } 
\end{equation}
which counts the number of $+1$ vertices minus the number of
$-1$ vertices, normalized by the total number of $+1$, $-1$ and $0$ vertices. 
The saddle point values of $\phi_{1,2,3}$ may also be substituted into $S_{\rm BEG}$ itself to obtain the free energy.

The explicit expressions of the saddle point values of $\phi_{1,2,3}$ are not particularly illuminating: there are two families of non-trivial solutions, one with $\phi_1 \ne \phi_2$ corresponding to a magnetized state and one with
$\phi_1 = \phi_2$ corresponding to an un-magnetized state. There is also a trivial solution with  $\phi_1 = \phi_2 = \phi_3=0$. The expression for the saddle point action $S_{sp}$ in the case of the magnetized solution(s) is, however, quite simple
\begin{equation}
\label{Ssp}
S_{sp}={1 \over 6} +{1 \over 12 \Delta^2}  - {a \over 2 \Delta} \pm {1 \over 12 \Delta^2}{\left( 1 - 4 \Delta a \right)^{3/2}}
\end{equation}
and makes it clear that we might expect  critical behaviour (at least) along the curve $4 \, \Delta_c a_c = 1$. The saddle point magnetization also confirms this since it is given along the various branches  by
\begin{equation}
\label{fullM}
M = \pm { \sqrt{\Delta (\Delta-2 a \pm 2 a \sqrt{1-4 \Delta a})} (2 \Delta-a \pm a \sqrt{1-4 \Delta a}) \over 2 \Delta^2-6 \Delta a \pm 4 \Delta a \sqrt{1-4 \Delta a}+1 \mp \sqrt{1-4 \Delta a}}
\end{equation}

To scan the phase diagram it is convenient to evaluate the magnetization $M$ at fixed $a$ as $\Delta$ is varied.
In Fig.\,(2) we have taken $a=0.1$ (i.e. a low temperature value) and plotted all the non-zero branches of the magnetization against $\Delta$.
\begin{figure}[h]
\begin{center}
\includegraphics[height=6cm]{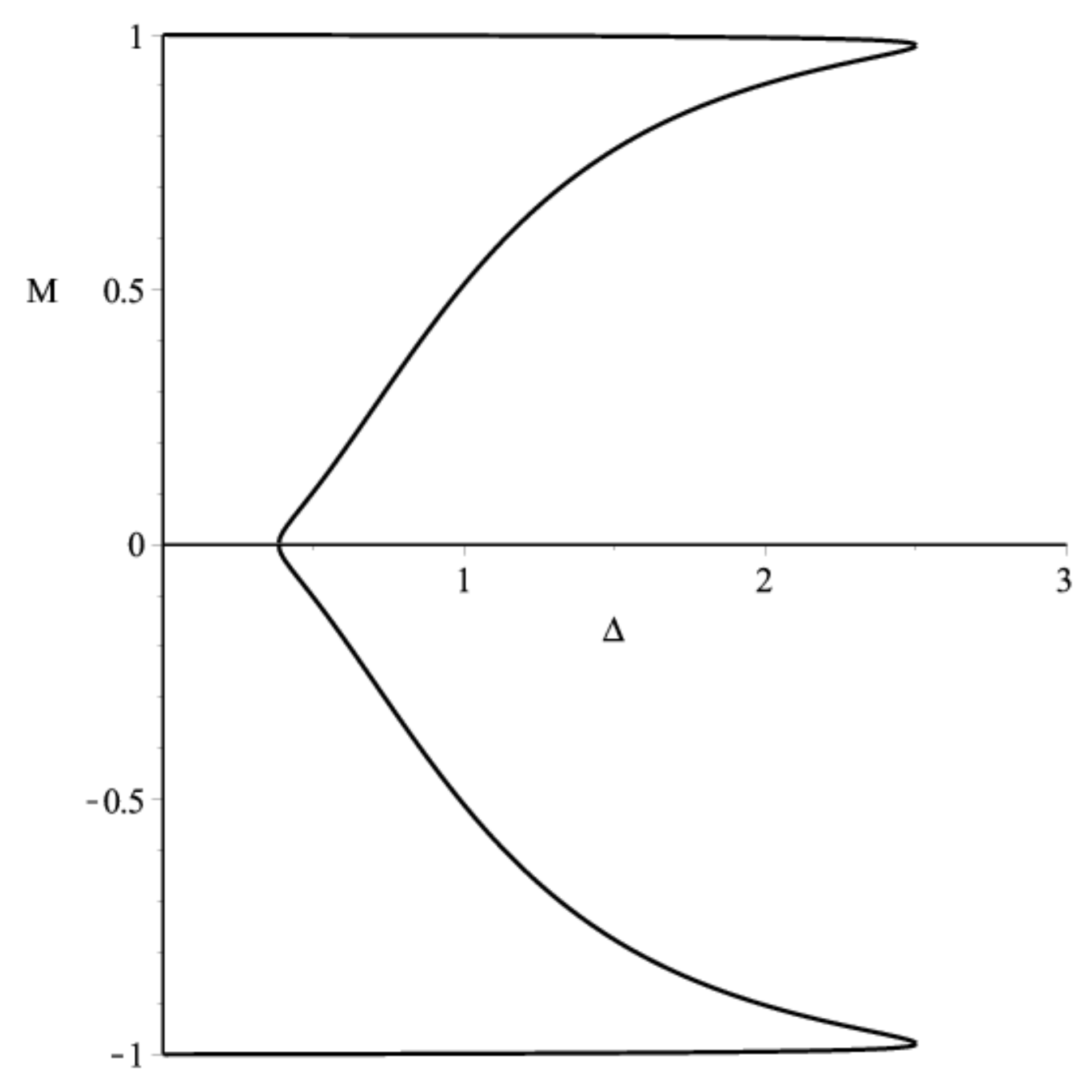}
\label{M_a_small} 
\caption{The magnetization plotted against $\Delta$ at $a=0.1$ (i.e. {\it low} temperature)}
\end{center}
\end{figure} 
The branching values at the tips of the prongs are at $\Delta_c=2.5$, confirming that $4 \Delta_c a_c =1$. As $\Delta$ is reduced at fixed $a$ (which also corresponds to the crystal field $\mu$ decreasing)  the magnetization will jump at $\Delta_c$ since the internal branches of the prong have a higher free energy, signalling a first order transition.

As $a$ is increased the tips of the prong move in towards the horizontal axis and by $a=0.37$ they have coalesced as shown in  Fig.\,(3)
\begin{figure}[h]
\begin{center}
\includegraphics[height=6cm]{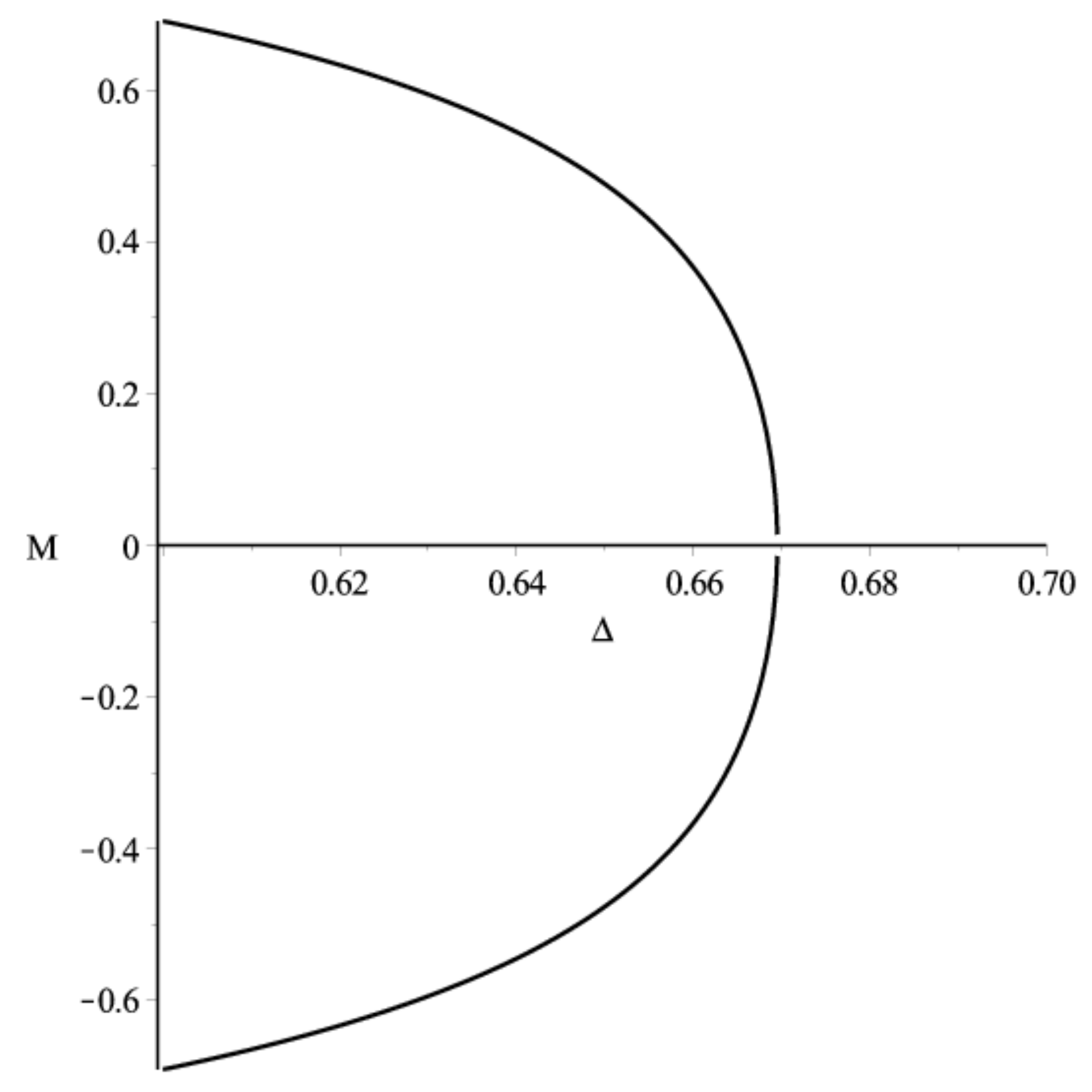}
\label{M_a_big} 
\caption{The non-zero magnetizations plotted against $\Delta$ at $a=0.37$ which lies just above the triple point value of $a_t=1 / \sqrt{8} \simeq 0.3535 \ldots$}
\end{center}
\end{figure} 
which gives a picture that is typical of a second order transition. We can pinpoint the triple point value $a_t$ where the prongs coalesce and the transition becomes second order by substituting the equation for the first order critical line $\Delta_c = 1 / 4 a_c$ into the saddle point magnetizations to get the magnetization at $a_c$, giving expressions of the form
\begin{equation}
M_c = \pm { \sqrt{ 1 - 8 a_c^2} ( 1 - 2 a_c^2) \over ( 4 a_c^2 -1 )}
\end{equation}
which show that $M_c=0$ at the triple point value, $a_t = 1 / \sqrt{8} \simeq 0.3535..$ with $\Delta_t = 1 / \sqrt{2} \simeq 0.7071..$.

On the second order transition line the critical value of $\Delta$ decreases as $a$ is increased, as can be seen from the full expression for  the non-zero branches of $M$ in equ.\,(\ref{fullM}) above
\begin{equation}
M = \pm { \sqrt{\Delta (\Delta-2 a + 2 a \sqrt{1-4 \Delta a})} (2 \Delta-a + a \sqrt{1-4 \Delta a}) \over 2 \Delta^2-6 \Delta a + 4 \Delta a \sqrt{1-4 \Delta a}+1 - \sqrt{1-4 \Delta a}}
\end{equation}
where the physical solutions for the second order branches are given by choosing the upper signs for the various square roots in equ.\,(\ref{fullM}).  As $a \rightarrow 1/2 $ the critical value of $\Delta$  tends to zero and no magnetized solution exists for the remainder of the physical range $1/2< a <1/ \sqrt{2}$.  

Remembering that increasing $a$ corresponds to increasing $T$ the general picture presented from the thin graph saddle point calculation is thus perfectly consistent with that observed from the Bragg-Williams mean-field approach shown in Fig.\,(1). For $0<a<a_t=1/\sqrt{8}$ first order transitions are observed as $\Delta$ is varied, whereas between $1/ \sqrt{8}<a<1/2$ second order transitions are observed with the critical value of $\Delta$ decreasing with increasing $a$ (i.e. increasing temperature). For $1/2<a<1/\sqrt{2}$ on the other hand, there is no transition.

Having obtained the phase diagram of the BEG model on thin $\phi^3$ graphs we are now in a position to re-interpret  it as the phase diagram of a ($2$,$r$)-state Potts model on the same graphs.
The key equation here is equ.(\ref{a3r}) which states the relation between the number of invisible states and the parameters in $S_{\rm BEG}$, namely $a^3 r = \Delta$. The key values to insert in this are those of the triple point where the transition changes from being second to first order, $a_t = 1 / \sqrt{8}$ and $\Delta_t = 1 / \sqrt{2}$, so we find that $r$ at this point has the surprisingly large value of  16.

In summary, on $3$-regular random graphs the ($2$,$r$)-state Potts model requires 17 or more invisible states to transmute the second order transition displayed by the $2$-state Potts (Ising) model on such graphs 
into a first order transition.

\section{The Wajnflasz-Pick and the standard Potts model on thin graphs}

Curiously, another mechanism for changing the order of a phase transition discussed by Tanaka and Tamura in \cite{TT4},
the Wajnflasz-Pick model \cite{WP,WP2}, bears  a close relation to the behaviour of the standard Potts model (with no invisible states) on thin graphs \cite{JP2}.
The Wajnflasz-Pick Hamiltonian is Ising-like, but there are $g_+$ positive spin states and $g_-$ 
negative spin states
\begin{equation}
{\cal H}_{\rm WP} = -J \sum_{\langle i,j \rangle} s_i s_j - h \sum_i s_i,
  \quad 
  s_i = \underbrace{+1, \cdots, +1}_{g_+}, \underbrace{-1,\cdots,-1}_{g_-} \; .
\end{equation}
This Hamiltonian can be transformed to a standard Ising Hamiltonian in a shifted temperature dependent external field \cite{WP2}
\begin{eqnarray}
 {\cal H}_{\rm WP} = -J \sum_{\langle i,j\rangle} \sigma_i \sigma_j
  - (h-\frac{T}{2}\log \frac{g_+}{g_-}) \sum_i \sigma_i,
  \quad
  \sigma_i = +1, -1 \; .
\end{eqnarray}
\begin{figure}[h]
\begin{center}
\includegraphics[height=6cm]{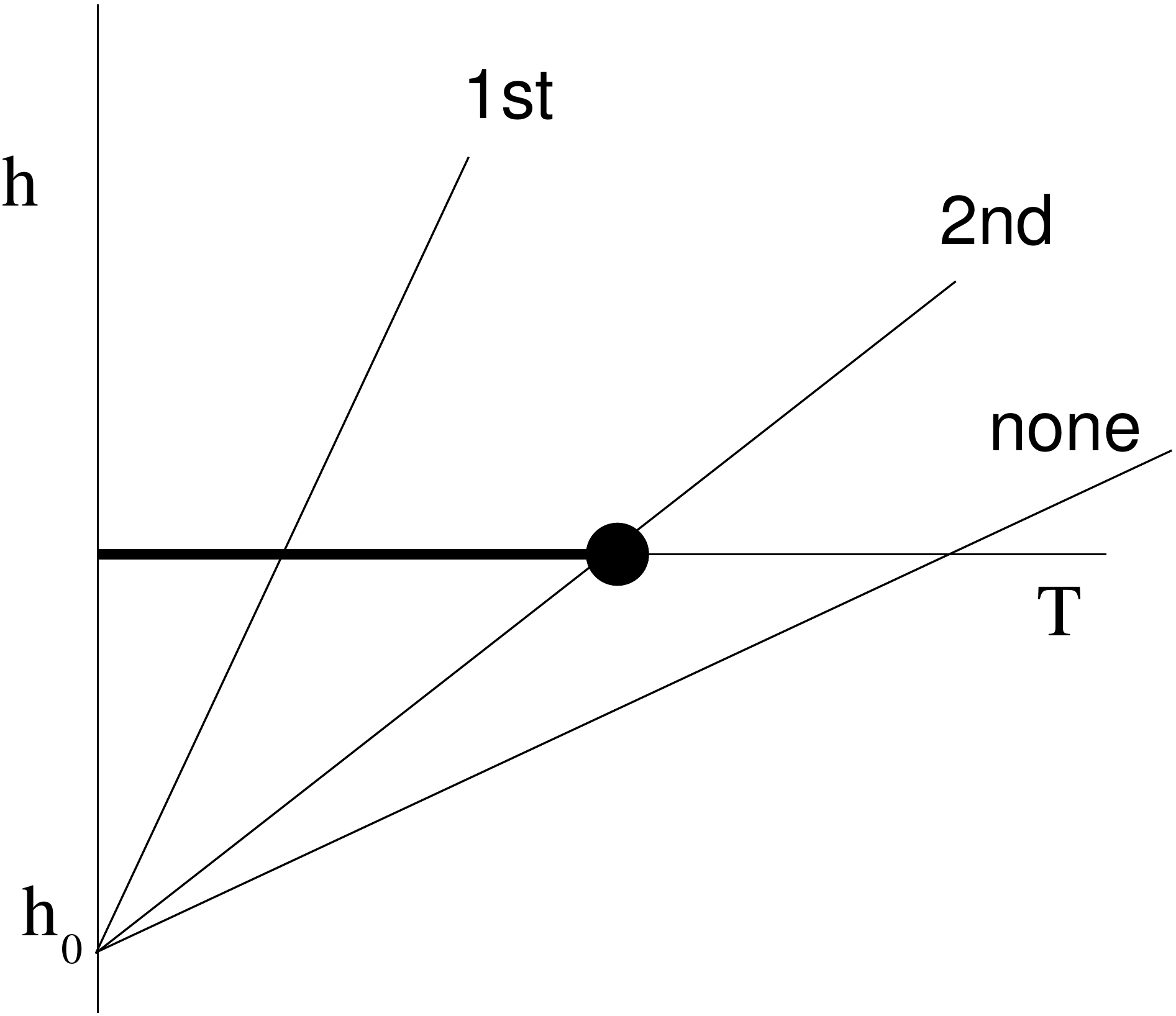}
\label{WP} 
\caption{The $h,T$ phase diagram for the Wajnflasz-Pick model. The trajectory of the system at fixed external field $h_0$ as $T$ is increased is plotted for three ratios
of $g_+$ to $g_-$.}
\end{center}
\end{figure} 
We can see in Fig.\,(4) that changing the temperature at a fixed external field $h_0$ will take the system along a line 
with no transition, a second order transition (if it goes through $T_c$) or a a first order transition depending on the ratio $g_+ / g_-$

We now consider the standard q-state Potts model
where the Hamiltonian is that given in equ.\,(\ref{Potts1})
\begin{equation*}
{\cal H}_q =   - \sum_{\left<ij\right>}  \delta_{\sigma_i, \sigma_j} \; .
\end{equation*}
On thin graphs the Potts models  display the mean-field behaviour of a continuous transition for $q=2$ and  first order transitions
for $q>2$. 
The model on $3$-regular thin graphs is described by the action of equ.\,(\ref{qstate}), also repeated here for convenience,
\begin{equation*}
S_q = { 1 \over 2 } \sum_{i=1}^{q} \phi_i^2 - c \sum_{i<j} \phi_i \phi_j -{1 \over 3} \sum_{i=1}^q \phi_i^3  \nonumber
\end{equation*}
and one finds a high temperature saddle-point solution of the form $\phi_i= 1 - (q-1)c, \forall i$
bifurcating to a broken symmetry solution $\phi_1= \ldots \phi_{q-1} \ne \phi_q$
at $c=1/(2 q - 1)$ where, as before, $c = 1/ ( \exp( 2 \beta)  + q-2)$. 
The magnetization may be defined as
\begin{equation}
m = { \phi_q^3 \over \left( \sum_{i=1}^{q} \phi_i^3 \right)}
\end{equation}
with a corresponding order parameter given by
\begin{equation}
M = { q \max ( m ) - 1 \over q -1 }.
\end{equation}
which is zero in the high temperature paramagnetic phase and tends
to one in the magnetised low temperature phase.

The graph of $m$, shown in in Fig.\,(5) for a $3$-state Potts model, is a skewed pitchfork but the phase transition takes place
neither at $O=(q-1-2\sqrt{q-1})/(q-1)(q-5)$ where the low temperature branches meet, nor at $P=1/(2q-1)$ where the high temperature, symmetric branch intersects the low temperature branch, but rather at a value $Q=(1-(q-1)^{-1/3})/(q-2)$ which lies between the two where the high and low temperature actions $S$ (i.e. free energies) are equal \cite{JP}. There is therefore  a jump in the magnetization 
of $\Delta M = (q-2)/(q-1)$ at this point between the high temperature value 
and that on the magnetized branch. When $q=2$ (the Ising case) the pitchfork is symmetric and the points $O,P,Q$ coincide to give a continuous transition
\begin{figure}[h]
\begin{center}
\includegraphics[height=6.5cm]{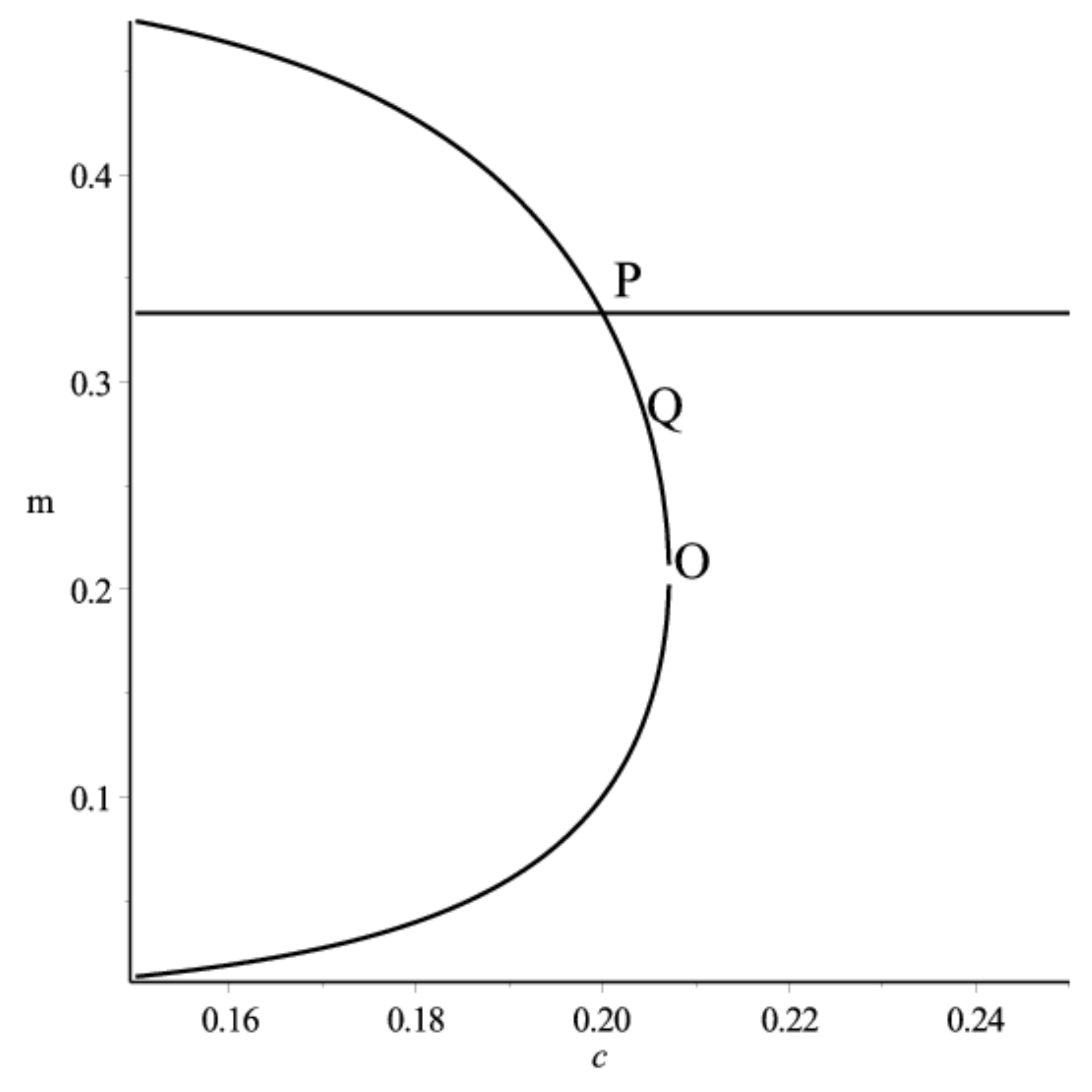}
\label{Phase} 
\caption{The magnetisation
$m$
for a $3$ state Potts model as calculated from
the saddle point solutions. 
Only the portion of the graph close to the transition point is shown for clarity.}
\end{center}
\end{figure} 

At first sight, this looks rather different to the picture presented by the Wajnflasz-Pick model, but we can recast the Potts model transition on thin graphs as that of an Ising model in a temperature dependent external field \cite{JP2} by noting that the symmetry breaking pattern is $q \rightarrow (1, q-1)$, so setting $\phi_1 = \phi$ and  $\phi_2=\phi_3= \ldots = \phi_q = \tilde \phi$ in $S_q$ to give
\begin{eqnarray}
{\tilde S}_q = {1 \over 2} ( q - 1) \left[ 1 - c ( q - 2) \right] \phi^2  - { \lambda \over 3} ( q -1) \phi^3
+ {1 \over 2} \tilde \phi^2  - {\lambda \over 3} \tilde \phi^3 - c ( q -1 ) \phi \tilde \phi.\nonumber\\
{}
\label{app1}
\end{eqnarray}
is sufficient to capture the transition.
The action of equ.\,(\ref{app1}) containing only ($\phi, \tilde \phi$)  may then be rescaled using 
\begin{eqnarray}
\phi \rightarrow { 1 \over \sqrt{( q - 1) ( 1 - c ( q - 2) )}} \phi ,
\label{rescal1}
\end{eqnarray}
to give an action equivalent to that of an Ising model in field
\begin{equation}
{\tilde S}_I = { 1 \over 2} ( \phi^2 + \tilde \phi^2 ) - \kappa
\phi \tilde \phi -{ \lambda v  \over 3 } \phi^3 
-{\lambda \over 3} \tilde \phi^3
\label{app2}
\end{equation}
where the new parameters $v$ and $\kappa$ are related to the original $c$ and $q$ via 
\begin{eqnarray}
\label{params}
v &=&   { 1 \over  (q-1)^{1/2} (1 - ( q - 2 ) c)^{3/2} } \nonumber \\
\kappa &=&  \sqrt{ c^2 ( q - 1) \over 1 - ( q - 2) c } \; .
\end{eqnarray}
The external field parameter $v$ in equs.\,(\ref{app2},\ref{params}) depends on the temperature $\beta$ via $c$, so the temperature driven first order transition of the standard Potts model on thin graphs appears in this formulation as the first order {\it field} driven transition of the Ising model. Substituting
the value of $c$ at the first order transition point, $Q=(1-(q-1)^{-1/3})/(q-2)$, into the first of equ.\,(\ref{params}) gives
$v=1$, the zero field point, so the picture of the transition in this formulation is similar to that presented by Fig.\,(4) where the first order transition occurs when the system crosses the first order line in the $h,T$ plane. If we take $q=2$, then $v=1$ for all $c$ and the action becomes that of the standard Ising model, which displays a second order transition.

\section{Discussion and Conclusions}
We have shown that it is possible to determine the phase diagram of the BEG model
with equal couplings  on $3$-regular ($\phi^3$) random graphs by considering the saddle point equations derived from an action which decorates the graphs with the appropriate statistical weights. The mean-field phase diagram thus derived shares the properties of other mean-field approaches and displays both first and second order transitions.

Using the equivalence of the BEG model and a ($2$,$r$)-state Potts model with $r$ ``invisible'' states implies that the position of the triple point in the BEG model can then be interpreted as showing that  a ($2$,$r$)-state Potts model  requires 17 or more invisible states in order to transmute the 
second order transition of the Ising model on  $\phi^3$  random graphs into a first order transition. Although the number of invisible states $r$ needed to force a first order transition is larger than the other examples explored by Tamura {\it et al}, the position of the triple point which determines the critical value of $r$ for such a transmutation is not a universal quantity and would be expected to be lattice dependent.  

As further evidence of such non-universality, we could take the $\phi$ in equ.\,(\ref{SBEG},\ref{SBEGeq}) to be $N \times N$ hermitean matrices and consider the $N \to \infty$ limit
of the integral rather than the $N \to 1$ limit which has effectively been considered here for the thin graphs. In this case we would be evaluating the partition function of the BEG model on an ensemble of {\it planar} random graphs, or if one preferred the  partition function of the BEG model coupled to 2D quantum gravity. This was done by Fukazawa {\it et al} \cite{3M1} for the equal coupling BEG model on $\phi^4$ planar graphs (i.e. planar, 4-regular random graphs), who found that the triple point coupling values on such graphs were given by numerical means as
approximately $a_t^2=0.06111$ and $\Delta_t=0.8329$. In this model the relation between the crystal field $\mu$ and the couplings $a$ and $\Delta$ is given by $a^4 \exp (\beta \mu) = \Delta$,
so taking $\mu = T \ln r$ as before we have $a^4 r = \Delta$ and a critical $r$ value of
approximately $223$. On such graphs a ($2$, $r$)-state Potts model would display a third order transition (the order of the Ising transition on such graphs) for up to 223 invisible states. The general effect of coupling spin models to planar random graphs  is to ``soften'' the order of the phase transition (the Ising transition becomes third order, for example), so many more invisible spins appear to be necessary to overcome this compared with, for instance, a square $2D$ lattice.

We also noted that the first order transition of the standard $q$-state Potts model on thin graphs could be thought of as a different sort of transmutation of the continuous Ising transition on such graphs, analogous to the mapping between the Wajnflasz-Pick model and the Ising model in an external field also discussed in \cite{TT4}.
We remark in closing that the BEG model has been studied on the Bethe
lattice using recursion relations in \cite{Anan}. It would be an interesting exercise to confirm that the saddle point equations used here were equivalent to the fixed point of these recursion relations, as is the case 
for other thin graph models which have been compared with their Bethe lattice counterparts \cite{JP2}.

\section{Acknowledgements}
R. P. K. C. M. Ranasinghe would like to thank the University of Sri Jayewardenepura for foreign leave and the Maxwell Institute for Mathematical
Sciences for hospitality.


\end{document}